\documentclass[                  
 	,draft
               nofootinbib,         
               ]{aipproc}
               
\layoutstyle{6x9}

\usepackage{amssymb,epsfig,amsmath,enumerate,undertilde}
\usepackage[usenames,dvipsnames]{color}

\begin{document}

\title{Initial data giving rise to naked singularities in spherically symmetric dust collapse}

\author{N\'estor Ortiz}{
address={Instituto de F\'{\i}sica y Matem\'aticas,
Universidad Michoacana de San Nicol\'as de Hidalgo\\
Edificio C-3, Ciudad Universitaria, 58040 Morelia, Michoac\'an, M\'exico.}
} 
\begin{abstract}
The phase space corresponding to a particular four-parameter family of initial data for the gravitational collapse of a spherically symmetric dust cloud is investigated. In a certain limit of the parameters, this family reproduces the case of homogenous mass density -constant mass distribution- and zero initial velocity, while in another limit, it generates a globally naked singularity. We show that for initial data characterizing flat density profiles, as well as large initial velocities, the probability of forming a globally naked singularity is low.
\end{abstract}
\maketitle
\section{Introduction}
\label{Sect:Intro}
In General Relativity, the {\it singularity theorems} \cite{HawkingEllis-Book} predict that a spacetime singularity forms when a sufficiently large mass is concentrated in a small region of an asymptotically flat spacetime, as for instance occurs in the complete gravitational collapse of a star. An open question is whether this singularity is {\it naked} (light rays emanating from it are visible by distant observers), or {\it hidden} (a black hole {\it censures} the singularity so that no information escaping from it can be detected outside of the event horizon). According to the weak cosmic censorship conjecture \cite{rP69}, the second must occur in the generic, realistic matter case \cite{Wald-Book}. In this work, we consider the simple model of a spherically symmetric dust cloud, whose collapse can lead to the formation of globally visible shell-focusing singularities \cite{dC84,rN86,nOoS11,dElS79} without fine-tuning the initial data, indicating that they are stable within this particular model. Although the model is restricted to spherical symmetry and a pressure-less fluid (and thus its matter content is not fully realistic), the formation of naked singularities is interesting by itself. In this article, we construct a four-parameter family of initial data for the spherical dust collapse. For certain values of the parameters, it gives a hidden singularity from constant mass density and zero initial velocity, while for other values, it gives a naked singularity. We construct the corresponding conformal diagrams using numerical techniques~\cite{nOoS11,nOoS10} and analyze the phase space of the parameters of the initial data.
\section{Tolman-Bondi dust collapse model}
Inside a self-gravitating, spherically symmetric collapsing dust cloud of finite radius $R_1$, in co-moving, synchronous coordinates $(\tau,R)$~\cite{MTW-Book}, the metric ${\bf g}$ and four-velocity ${\bf u}$ are

\begin{equation}
{\bf g} = -d\tau^2 + \frac{r'(\tau,R)^2}{1 + 2E(R)}\; dR^2
 + r(\tau,R)^2(d\vartheta^2 + \sin^2\vartheta\, d\varphi^2),
\quad {\bf u} = \frac{\partial}{\partial\tau}\;,\label{Eq:MetricSol}
\end{equation}
where, for given $\tau$, $r(\tau,R)$ is the areal radius of a matter shell of initial radius $R$, this means that the coordinate $R$ is such that $r(0,R) = R$. Our assumptions below exclude shell crossing singularities; therefore, each matter shell is uniquely characterized by $R$. Here and in the following, a prime denotes partial derivation with respect to $R$ and a dot with respect to $\tau$. From Einstein's equations, the time evolution of $r$ is determined by
 
\begin{equation}
\frac{1}{2} \dot{r}(\tau,R)^2 + V(r(\tau,R), R) = E(R),\qquad
V(r,R) := -\frac{m(R)}{r},
\label{Eq:1DMechanical}
\end{equation}
where $m(R)$ is the Misner-Sharp mass function~\cite{cMdS64} and $E(R) = v_0(R)^2/2 - m(R)/R$ is the total energy for each shell $R$, given in terms of the initial data, which consist of the profiles for the initial velocity $v_0(R)$ and the initial density $\rho_0(R)$. Now, it is convenient to introduce the new functions
\begin{displaymath}
c(R) := \frac{2m(R)}{R^3},\qquad
q(R) := \sqrt{E(R)/V(R,R)} = \sqrt{1 - \frac{R v_0(R)^2}{2m(R)}}.
\end{displaymath}
Note that $c(R)$ is directly proportional to the mean density and $q(R)^2$ is the ratio between the total and initial potential energy. The initial data must satisfy specific conditions such as regularity, smoothness, etc. (see Ref.~\cite{nOoS11} for details and explanations), which imply
\begin{equation}
c(R) > 0, \quad c'(R)\leq 0, \quad 0 < q(R) < 1, \quad q'(R)\geq 0, \label{Eq:cqConditions}
\end{equation}
as well as $q'(R)/R > 0$ whenever $c'(R)/R = 0$, for $R\geq 0$. This guaranties \cite{nOoS11} the existence of light rays escaping from the central singularity, making it visible, at least to local observers.
These assumptions do not cover the Oppenheimer-Snyder case~\cite{jOhS39}, ($E=0$ and constant mass density profile), however, it can be recovered by approximation from data satisfying our conditions.
\section{Initial data and conformal diagrams}
A particular choice for the functions $c$ and $q$ is the four-parameter family
\begin{equation}
c(R) := c_0 \left[ 1 - \frac{3}{3+2n}\left( \frac{R}{R_1}\right)^{2n} \right],\qquad
q(R) := q_0 + q_1\left( \frac{R}{R_1} \right)^2,\qquad
0 \leq R \leq R_1,
\label{Eq:cqChoice}
\end{equation}
where $R_1 > 0$ is the initial radius of the cloud (for $R > R_1$ we have Schwarzschild spacetime). The parameters are subject to $n \geq 1$, $0 < R_1^2 c_0 < (3+2n)/2n$, $0 < q_0 < 1$, and $0 < q_1 < 1-q_0$, which guarantee the satisfaction of conditions (\ref{Eq:cqConditions}) on the interval $[0,R_1]$. The corresponding initial density and velocity profiles are $(\rho_0,v_0) = ( (R^3 c)'/(8\pi G R^2), -R\sqrt{(1-q^2)c})$. Explicitly, the resulting density profile is
\begin{displaymath}
\rho_0(R) = \frac{3c_0}{8\pi G}\left[ 1 - \left(\frac{R}{R_1}\right)^{2n} \right],\qquad
0 \leq R \leq R_1.
\end{displaymath}
The important feature of this choice is that $\rho_0$ becomes constant as $n \to \infty$, moreover, the initial velocity becomes zero as  $q_0 \to 1$ and $q_1 \to 0$ (Note that the Oppenheimer-Snyder case is recovered when both $q_0, q_1 \to 0$). Figures~\ref{Fig:naked}~and~{\ref{Fig:hidden} (left panel) are examples of numerically generated conformal diagrams -- in coordinates $(T,X)$, in which the radial part of the metric~(\ref{Eq:MetricSol}) takes the form $\Omega(T,X)^2\left( -dT^2 + dX^2 \right)$, with a conformal factor $\Omega(T,X) > 0$ --, with initial data given by (\ref{Eq:cqChoice}) with $n=41$ (nearly constant density), $q_0 = 0.98$ and $q_1 = 0.01$ (nearly zero initial velocity) for both diagrams. The only parameter that has a different value in Figs.~\ref{Fig:naked} and~\ref{Fig:hidden} is  $c_0$. In Fig.~\ref{Fig:naked}, $c_0=0.063$, so the central density $\rho_0 = 3c_0/(8\pi G)$ is sufficiently small to generate a globally naked singularity, as predicted by theorem~2 in Ref.~\cite{nOoS11}. In Fig.~\ref{Fig:hidden}, $c_0$ has the limit value $c_0 = 1.0$, which gives a hidden singularity. Its right panel shows the exact diagram for homogeneous -constant- density and zero initial velocity. The differences between the two panels are due to the implementation of different boundary conditions for their construction. Note that, while Fig.~\ref{Fig:naked} shows the complete spacetime, including the outside of the collapsing cloud, for simplicity Fig.~\ref{Fig:hidden} only shows the interior spacetime of the cloud.
\begin{figure}[h!]
\includegraphics[height=.35\textheight]{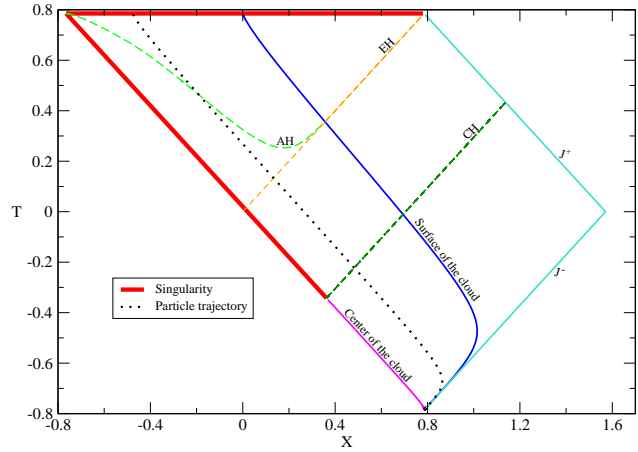}
\caption{Conformal diagram for the Tolman-Bondi dust collapse with initial data given by Eq.~(\ref{Eq:cqChoice}), with $q_0 = 0.98$, $q_1=0.01$, $n=41.0$ and $c_0 = 0.063$. The notation "AH", "EH" and "CH" refer to the apparent, event and Cauchy horizons, respectively. In this case, a portion of the singularity is globally naked. The dotted line corresponds to the dust particle trajectory with initial areal radius $R_0 = 0.95R_1$. The diagram was generated following the numerical algorithm described in Ref.~\cite{nOoS11}.}
\label{Fig:naked}
\end{figure}
\begin{figure}[h!]
\includegraphics[height=.295\textheight]{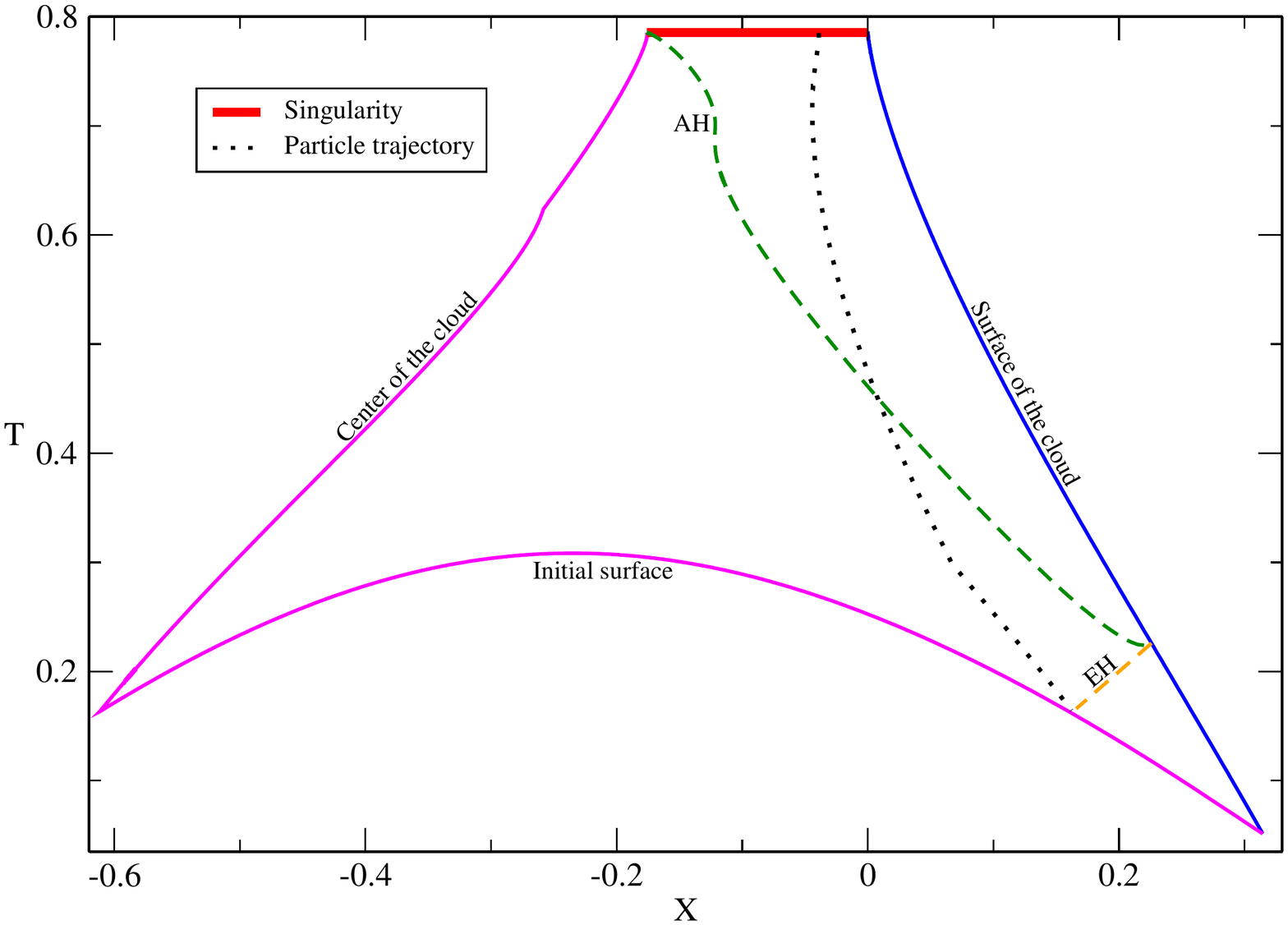}
\hspace{-1.15cm}
\includegraphics[height=.295\textheight]{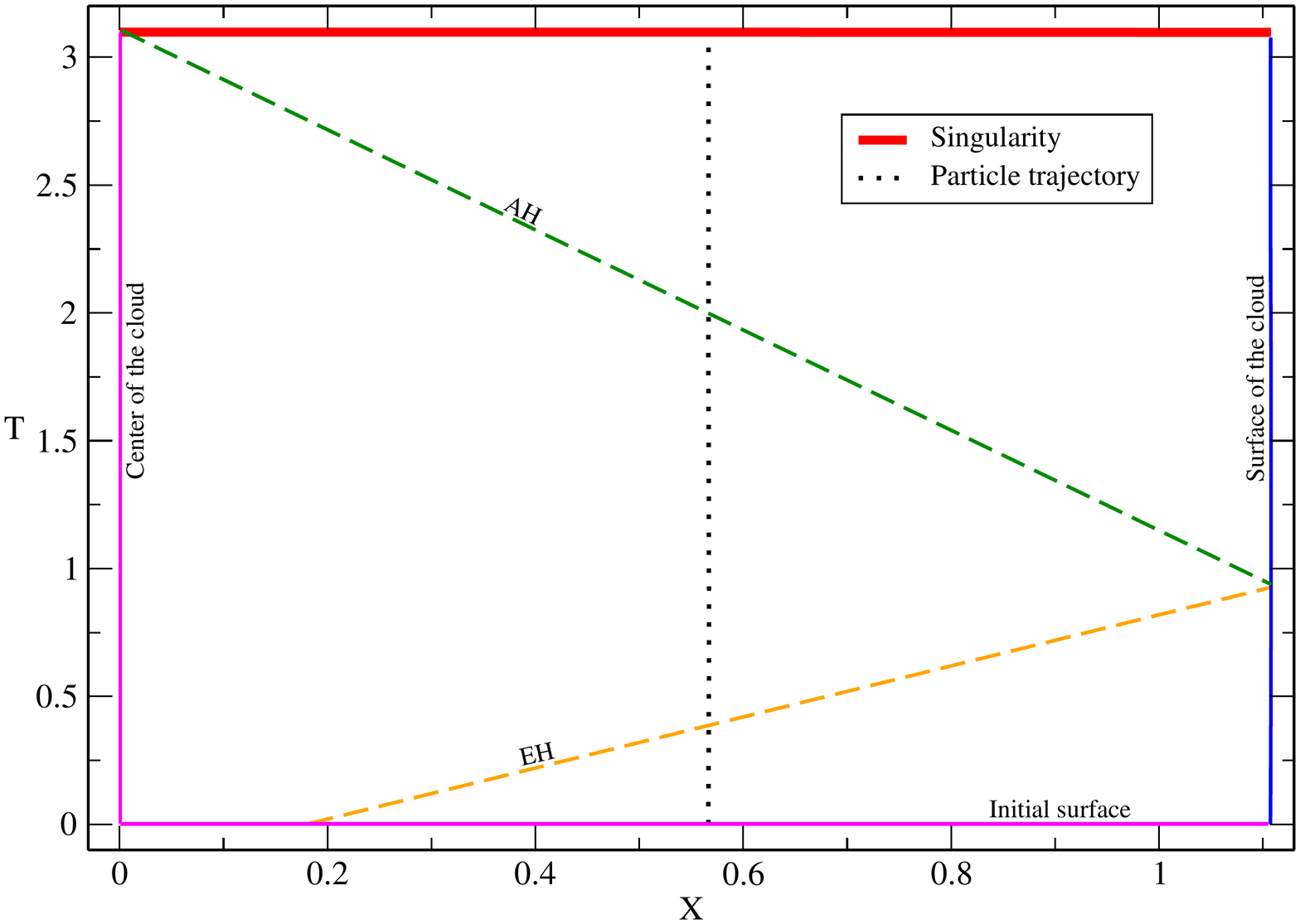}
\caption{{\bf Left panel}: Conformal diagram for the Tolman-Bondi dust collapse with initial data given by Eq.~(\ref{Eq:cqChoice}), with the same parameter choice as in Fig.~\ref{Fig:naked}, except for $c_0=1.0$. This is a hidden singularity inside a black hole. This singularity has a null portion as in Fig.~\ref{Fig:naked}, but here it is too small to be seen without a zoom. This is an approximation to the scenario of uniform density and zero initial velocity. {\bf Right panel}: Exact conformal diagram for the case of uniform density and zero initial velocity, see Ref.~\cite{nOoS11}.}
\label{Fig:hidden}
\end{figure}
Also note that in Fig.~\ref{Fig:naked} the apparent horizon in spacelike everywhere inside the cloud, while in Fig.~\ref{Fig:hidden} it has a timelike portion.
\section{Phase space}
\label{Sect:phase_space}
The phase space corresponding to the four-parameter family of initial data given in Eq.~(\ref{Eq:cqChoice}) is studied in this section. The left panel of Fig.~\ref{Fig:phase_space} shows a subset of this space for fixed $q$. The critical -continuous- line in the $n$-$c_0$-plane divides the regions corresponding to initial data giving rise to black holes and globally naked singularities (shaded region). It results that flat density profiles (large $n$) favor the formation of black holes. It is also inferred that the critical value for $c_0$ becomes asymptotically constant for large $n$. On the other hand, the dashed line corresponds to the upper bound for $c_0$ obtained from theorem~2 in Ref.~\cite{nOoS11}. 
Note that although this bound is not optimal, it describes the correct qualitative behavior for the critical line. The right panel of Fig.~\ref{Fig:phase_space} shows the critical line in the $n$-$c_0$-plane for different values of $q_0$ and $q_1=0.01$.
Since $1 - q^2$ is the ratio between the initial kinetic and potential energies, this figure shows that large initial velocities in the negative radial direction favor the formation of black holes. All of this is consistent with the conclusions obtained in section~5.2 of Ref.~\cite{nOoS11} using a different family of initial data. Nevertheless, an important new feature obtained from Fig.~\ref{Fig:phase_space} is that, for large initial velocities, the region for naked singularities becomes small for $n \geq 2$, which means that it becomes very difficult to obtain a naked singularity in this case.
\begin{figure}[h!]
\includegraphics[height=.287\textheight]{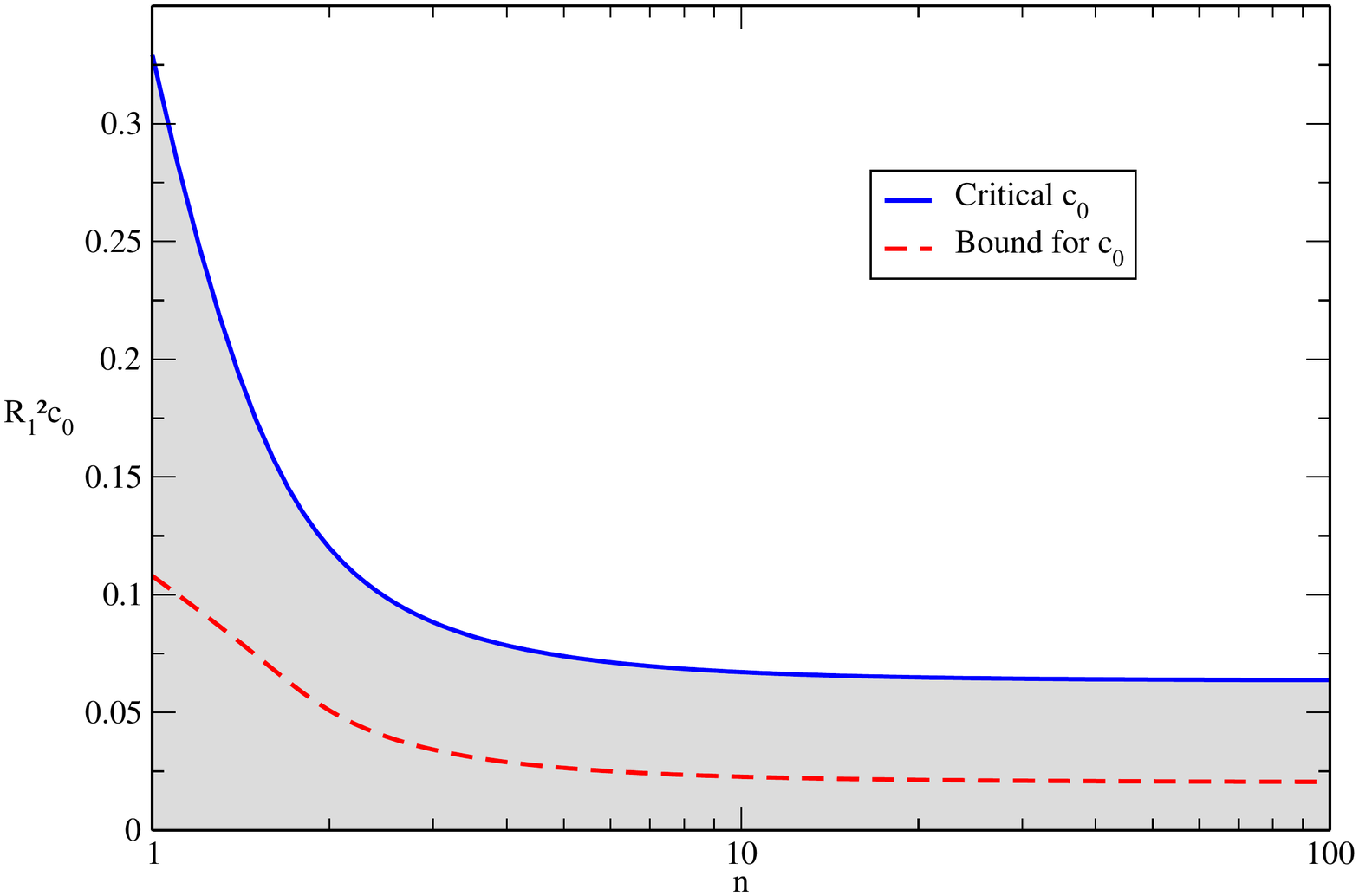}\label{Fig:phase_space}
\hspace{-1.23cm}
\caption{{\bf Left panel}: A cut of the parameter space of the four-parameter family of initial data~(\ref{Eq:cqChoice}), corresponding to the subset with fixed $q_0 = 0.98$ and $q_1=0.01$. The shaded region represents initial data giving rise to globally visible singularities. The dashed line describes the upper bound for $c_0$ given from theorem~2 in Ref.~\cite{nOoS11}. {\bf Right panel}: The critical line, dividing the regions of initial data giving rise to black holes and globally naked singularities, respectively, for fix $q_1=0.01$, and different values of $q_0$. Note that the region for naked singularities becomes very small for large initial velocities and flat density profiles.}
\includegraphics[height=.287\textheight]{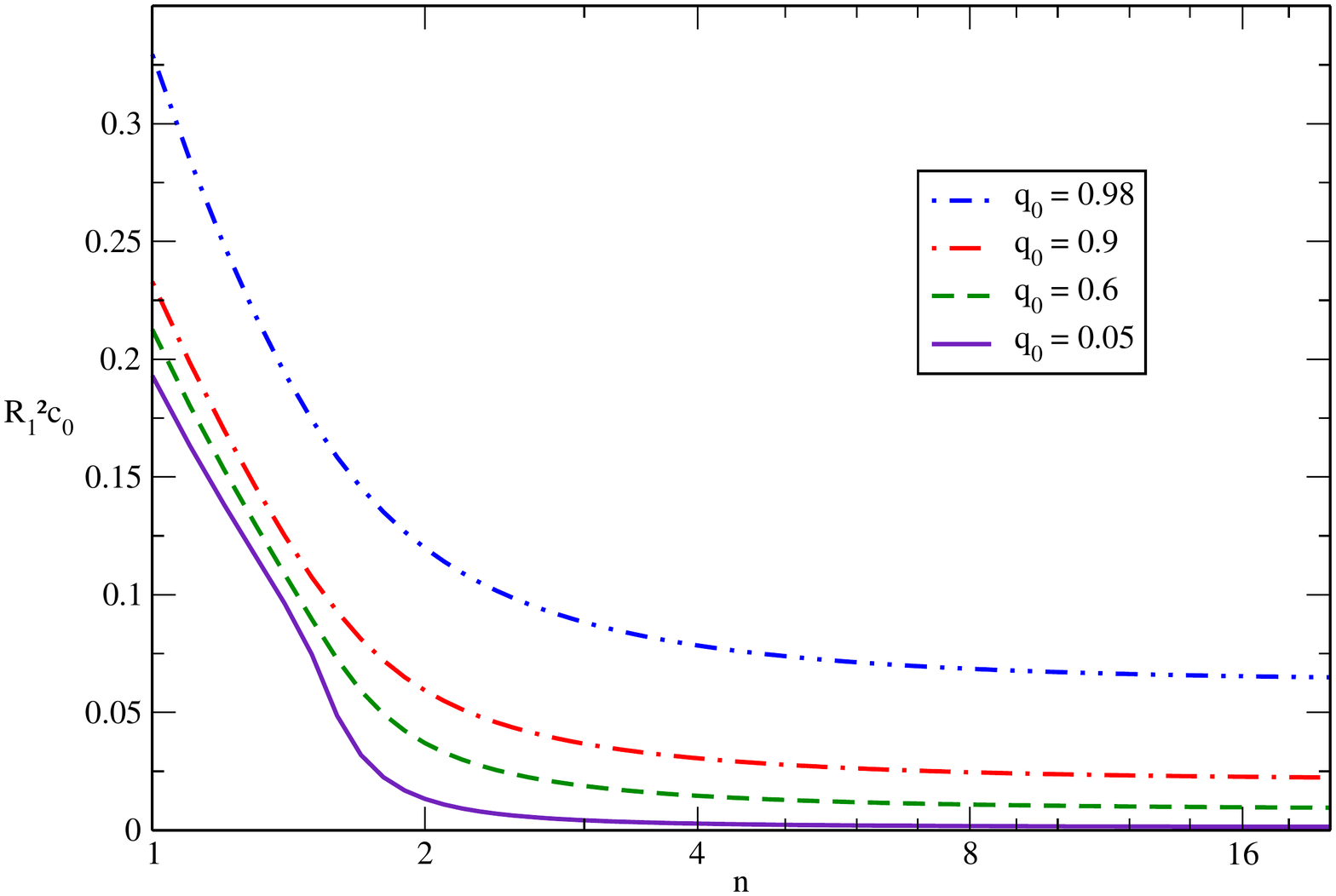}
\end{figure}
\section{Conclusions}
\label{Sect:Conclusions}
The Tolman-Bondi model for the gravitational collapse of a spherically symmetric dust cloud with fairly general initial density and velocity profiles was considered. A four-parameter family of initial data was constructed such that in a certain limit of the parameters, it arbitrarily approaches the case of homogenous -constant- density with zero initial velocity, while in another limit, a globally naked singularity is obtained. Some important remarks are: $(i)$ With an arbitrarily flat density profile, it is always possible to obtain a globally naked singularity for small central density. $(ii)$ Just as in the case of constant density and zero initial velocity, a portion of the apparent horizon may be timelike. $(iii)$ Flat density profiles favor the formation of black holes. $(iv)$ Large initial velocities favor the formation of black holes. $(v)$ For flat density profiles and large initial velocities, the probability of obtaining a naked singularity is very low. Since at the late stage of a collapse the velocity is high, this means that for a sufficiently flat profile it is difficult to obtain a naked singularity. 
\begin{theacknowledgments}
I am glad to thank Olivier Sarbach and Thomas Zannias for comments and discussions.
\end{theacknowledgments}
\bibliographystyle{aipproc}  
\bibliography{../References/refs_collapse}

\begin{thebibliography}{11}
\expandafter\ifx\csname natexlab\endcsname\relax\def\natexlab#1{#1}\fi
\providecommand{\enquote}[1]{``#1''}
\expandafter\ifx\csname url\endcsname\relax
  \def\url#1{\texttt{#1}}\fi
\expandafter\ifx\csname urlprefix\endcsname\relax\def\urlprefix{URL }\fi
\providecommand{\eprint}[2][]{\url{#2}}

\bibitem[Hawking and Ellis(1973)]{HawkingEllis-Book}
S.~Hawking, and G.~Ellis, \emph{The Large Scale Structure of Space Time},
  Cambridge University Press, Cambridge, 1973.

\bibitem[Penrose(1969)]{rP69}
R.~Penrose, \emph{Riv. del Nuovo Cimento} \textbf{1}, 252--276 (1969).

\bibitem[Wald(1984)]{Wald-Book}
R.~Wald, \emph{General Relativity}, The University of Chicago Press, Chicago,
  London, 1984.

\bibitem[Christodoulou(1984)]{dC84}
D.~Christodoulou, \emph{Comm. Math. Phys.} \textbf{93}, 171--195 (1984).

\bibitem[Newman(1986)]{rN86}
R.~Newman, \emph{Class. Quantum Grav.} \textbf{3}, 527--539 (1986).

\bibitem[Ortiz and Sarbach(2011)]{nOoS11}
N.~Ortiz, and O.~Sarbach, \emph{Class. Quantum Grav.} \textbf{28}, 235001
  (27pp) (2011).

\bibitem[Eardley and Smarr(1979)]{dElS79}
D.~Eardley, and L.~Smarr, \emph{Phys. Rev. D} \textbf{19}, 2239--2259 (1979).

\bibitem[Ortiz and Sarbach(2010)]{nOoS10}
N.~Ortiz, and O.~Sarbach, \emph{AIP Conf. Proc.} \textbf{1256}, 349--356
  (2010), proceedings of VIII Mexican School on Gravitation and Mathematical
  Physics.

\bibitem[Misner et~al.(1973)]{MTW-Book}
C.~Misner, K.~Thorne, and J.~Wheeler, \emph{Gravitation}, W. H. Freeman, 1973.

\bibitem[Misner and Sharp(1964)]{cMdS64}
C.~Misner, and D.~Sharp, \emph{Phys. Rev.} \textbf{136}, B571--B576 (1964).

\bibitem[Oppenheimer and Snyder(1939)]{jOhS39}
J.~Oppenheimer, and H.~Snyder, \emph{Phys. Rev.} \textbf{56}, 455--459 (1939).

\end{thebibliography}
\end{document}